\documentclass[aip,reprint]{revtex4-1}

\pdfoutput=1 

\usepackage{graphicx}
\usepackage{dcolumn}
\usepackage{bm}
\usepackage[mathlines]{lineno}

\usepackage[utf8]{inputenc}
\usepackage[T1]{fontenc}
\usepackage{mathptmx}
\usepackage{etoolbox}
\usepackage{hyperref}
\usepackage[caption=false]{subfig}
\usepackage{amsmath}
\usepackage{siunitx,upgreek}
\usepackage{xcolor,cancel}

\makeatletter
\def\@email#1#2{%
 \endgroup
 \patchcmd{\titleblock@produce}
  {\frontmatter@RRAPformat}
  {\frontmatter@RRAPformat{\produce@RRAP{*#1\href{mailto:#2}{#2}}}\frontmatter@RRAPformat}
  {}{}
}%
\makeatother

\begin{document}
\preprint{AIP/123-QED}

\title{Efficient Quantum Frequency Conversion of Ultra-Violet Single Photons from a Trapped Ytterbium Ion}

\author{Seungwoo Yu}
\affiliation{ 
Department of Computer Science and Engineering, Seoul National University, Seoul 08826, Republic of Korea
}
\affiliation{
Automation and System Research Institute, Seoul National University, Seoul 08826, Republic of Korea}
\affiliation{
Inter-University Semiconductor Research Center, Seoul National University, Seoul 08826, Republic of Korea}
\affiliation{
NextQuantum, Seoul National University, Seoul 08826, Republic of Korea}
\author{Kyungmin Lee}
\affiliation{ 
Department of Computer Science and Engineering, Seoul National University, Seoul 08826, Republic of Korea
}
\affiliation{
Automation and System Research Institute, Seoul National University, Seoul 08826, Republic of Korea}
\affiliation{
NextQuantum, Seoul National University, Seoul 08826, Republic of Korea}
\author{Sumin Park}
\affiliation{
Automation and System Research Institute, Seoul National University, Seoul 08826, Republic of Korea}
\affiliation{
NextQuantum, Seoul National University, Seoul 08826, Republic of Korea}
\affiliation{ 
Department of Physics and Astronomy, Seoul National University, Seoul 08826, Republic of Korea
}

\author{Kyunghye Kim}
\affiliation{ 
Department of Computer Science and Engineering, Seoul National University, Seoul 08826, Republic of Korea
}
\affiliation{
Automation and System Research Institute, Seoul National University, Seoul 08826, Republic of Korea}
\affiliation{
NextQuantum, Seoul National University, Seoul 08826, Republic of Korea}

\author{Junhong Goo}
\altaffiliation{Currently working at Samsung Display Co., Ltd.}
\affiliation{ 
Department of Computer Science and Engineering, Seoul National University, Seoul 08826, Republic of Korea
}
\affiliation{
Automation and System Research Institute, Seoul National University, Seoul 08826, Republic of Korea}
\author{Jeonghyun Park}
\affiliation{ 
Department of Computer Science and Engineering, Seoul National University, Seoul 08826, Republic of Korea
}
\affiliation{
Automation and System Research Institute, Seoul National University, Seoul 08826, Republic of Korea}
\affiliation{
NextQuantum, Seoul National University, Seoul 08826, Republic of Korea}
\author{Taehyun Kim*}
\email[Author to whom correspondence should be addressed:]{taehyun@snu.ac.kr}
\affiliation{ 
Department of Computer Science and Engineering, Seoul National University, Seoul 08826, Republic of Korea
}
\affiliation{
Automation and System Research Institute, Seoul National University, Seoul 08826, Republic of Korea}
\affiliation{
Inter-University Semiconductor Research Center, Seoul National University, Seoul 08826, Republic of Korea}
\affiliation{
NextQuantum, Seoul National University, Seoul 08826, Republic of Korea}
\affiliation{
Institute of Computer Technology, Seoul National University, Seoul 08826, Republic of Korea}
\affiliation{
Institute of Applied Physics, Seoul National University, Seoul 08826, Republic of Korea}

\date{\today}

\begin{abstract}
Ion trap system is a leading candidate for quantum network privileged by its long coherence time, high-fidelity gate operations, and the ion-photon entanglement that generates an ideal pair of a stationary memory qubit and a flying communication qubit. Rapid developments in nonlinear quantum frequency conversion techniques have enhanced the potential for constructing a trapped ion quantum network via optical fiber connections. The generation of long-distance entanglement has been demonstrated with ions such as Ca$^{+}$ and Ba$^{+}$, which emit photons in visible or near-infrared range naturally. On the other hand, as the qubit-native photons reside in ultra-violet (UV) spectrum, the Yb$^{+}$ ion has not been considered as a strong competitor for telecommunication qubits despite extensive research on it. Here, we demonstrate an efficient difference-frequency conversion of UV photons, emitted from a trapped Yb$^{+}$ ion, into a visible range. We provide experimental evidence that confirms the converted photons are radiated from the Yb$^{+}$ ion. Our results provide a crucial step toward realizing a long-distance trapped ion quantum network based on Yb$^{+}$ ions through quantum frequency conversion.
\end{abstract}
\maketitle

Trapped ion systems are promising platforms in quantum information science, offering long coherence times \cite{wang2021single} and high-fidelity gate operations through a phononic quantum information bus that leverages Coulomb interactions.\cite{srinivas2021high, gaebler2016high, PhysRevLett.127.130505} While the phonon interaction allows for precise manipulation of qubit states,\cite{PhysRevLett.127.130505} its implementation has been confined to small-scale systems within a single chain of ions so far. \cite{PhysRevX.13.041052, arXiv.2308.05071, arXiv.2407.07694} Photon channels have been used to generate entanglement between two distant ion qubits in separate traps,\cite{Nature449.7158.68, PhysRevLett.130.050803} and to realize a quantum teleportation protocol,\cite{Science323.5913.486} where ion-photon interaction offered an ideal entangled pair\cite{Nature428.6979.153} of a stationary memory qubit and a rapidly propagating communication qubit.\cite{FortschrPhys.57.1133, RevModPhys.82.1209, PhysRevA.84.063423} Combined with the phonon channel, photonic interconnects provide a reliable approach for realizing scalable quantum networks \cite{NatPhys11.1.37, FortschrPhys.57.1133} and large-scale quantum processors \cite{PhysRevA.89.022317} based on trapped ions. However, the significant attenuation of optical fibers has limited the distance of trapped ion quantum networks since the wavelengths of photons emitted by ion qubits are much shorter than the fiber-optic communication bands.\cite{Nature449.7158.68, Science323.5913.486} 

With recent advances in nonlinear frequency conversion techniques, photons in near-infrared (NIR) range emitted by trapped $\text{Ca}^{+}$ ion have been converted to telecom band for long-distance transmission,\cite{PhysRevLett120.20.203601} and high fidelity ion-photon entanglement was observed.\cite{npjQunatumInf5.1.72, NatComm9.1.1998} Based on these developments, a telecom-wavelength quantum repeater node for the generation of ion-ion entanglements over 50 km was realized through quantum frequency conversion (QFC).\cite{PhysRevLett.130.213601} Recent developments of $\text{Ba}^{+}$ ion have enabled a conversion of visible photons from $\text{Ba}^{+}$ ions to the NIR range,\cite{PhysRevApplied.11.014044} demonstrating ion-photon entanglement.\cite{PhysRevA.106.042441} By adopting 2-stage QFC setup, photons in visible range also have been converted to telecom band.\cite{ApplPhysLett.119.8.084001, ACSPhotonics.10.8.2861}

The ytterbium ion is one of the most popular and long-standing candidates for quantum computation \cite{PhysRevLett.104.140501, PhysRevLett.109.020503} and quantum communication \cite{Nature428.6979.153, NatPhys.3.8, Nature449.7158.68, Science323.5913.486, PhysRevLett.102.250502} based on trapped ion system due to its simple hyperfine energy level structure.\cite{PhysRevA.76.052314, RevModPhys.75.281} While the $\text{Yb}^{+}$ ion is widely used in the state-of-the-art trapped ion quantum processors,\cite{PhysRevX.13.041052, arXiv.2308.05071} it has not been well considered for the realization of quantum network over long distance due to its qubit-native photons in ultra-violet (UV) range. Therefore, ytterbium ions are used to generate entanglement\cite{Nature449.7158.68, PhysRevLett.102.250502} and demonstrate quantum teleportation \cite{Science323.5913.486} over only short distances.

\begin{figure}[t]
\includegraphics[width=\linewidth]{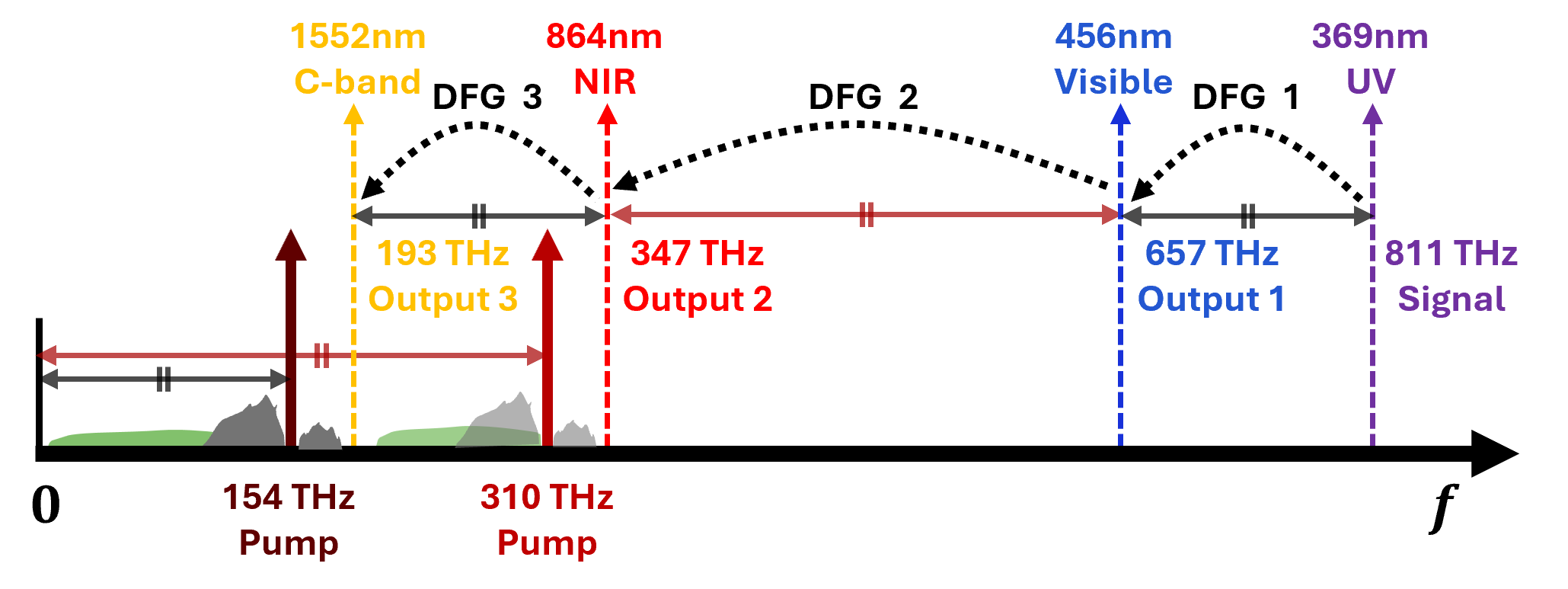}
\caption{A three-stage QFC scheme to convert 369-nm photons emitted by Yb$^{+}$ ions into 1552 nm. In this process, a UV photon is sequentially converted through DFGs into photons in the visible spectrum, then the NIR, and finally into the telecom C-band. The gray regions and green regions indicate the noises from SRS and SPDC, respectively. Note that DFG 2 is not affected by SPDC noise, as the 310-THz pump has a longer wavelength than the output photons at 864 nm. We conducted the initial stage (DFG 1), which presents the greatest challenge and is critical for achieving efficient conversion of the UV photons from Yb$^{+}$ ions. 
}
\label{figure1:convstage}
\end{figure}

\begin{figure*}[t]
\includegraphics[width=\linewidth]{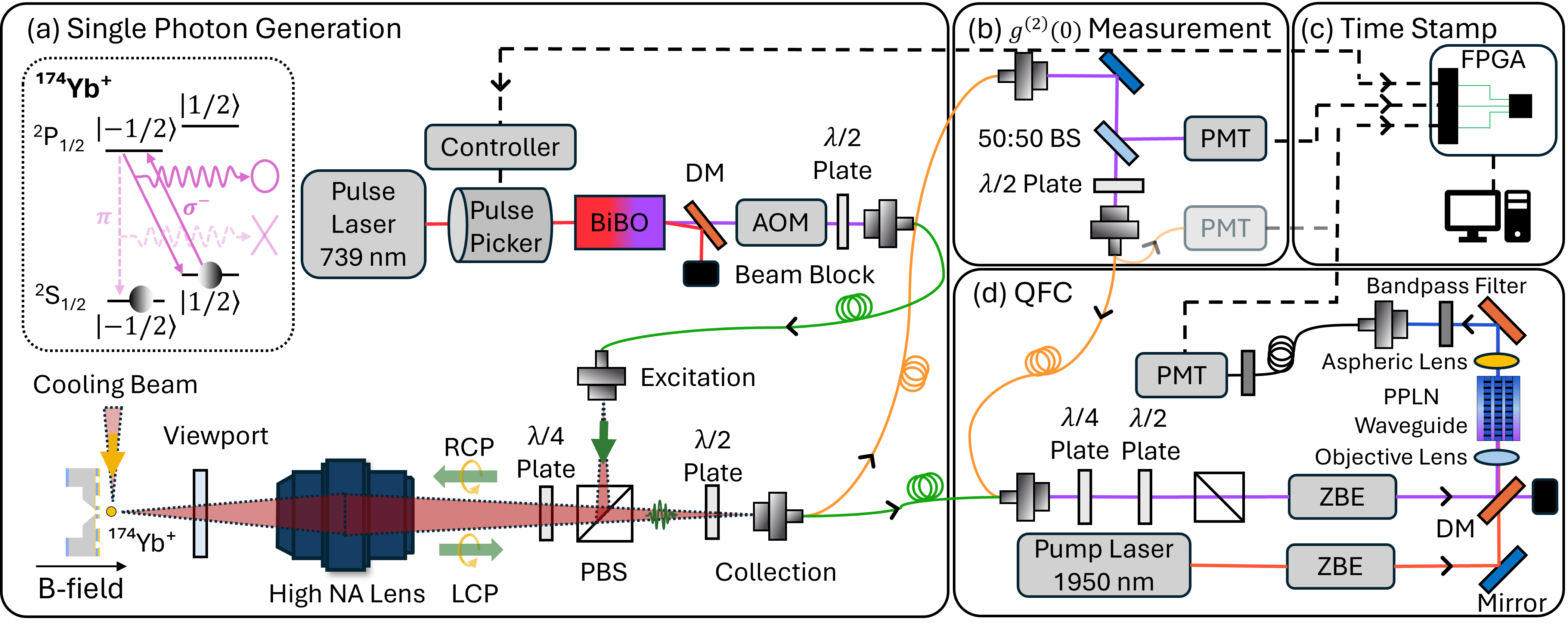}
\caption{Overall schematics of our experimental setup. The orange fiber lines are used to measure second-order correlation. The lifetime of the $^{174}$Yb$^{+}$ ion is measured with the green fiber lines. All the used fibers are PM fibers. (a) Ion trap and pulse laser setup are used to generate single photons. Single photons can be generated through the Doppler cooling process with a 369-nm cooling beam, or by the pulse laser that excites the $^{174}$Yb$^{+}$ ion. In both cases, only photons from $\sigma^{-}$-transitions could be detected through spatial filtering.\cite{PhysRevA.84.063423} The pulse picker sends a single laser pulse to the ion when it receives a transistor-transistor logic (TTL) signal sent from the FPGA. (b) A HBT-type setup is used to measure the second order correlation. The intensity correlation was measured using the photons from the PM fiber, either without any frequency conversion or by sending photons in one arm through the frequency conversion setup. (c) FPGA-based time stamper recording the arrival times of photons detected by the PMT. (d) QFC setup for difference-frequency conversion of 369-nm photons to 456 nm. The polarization and collimation of the laser light from the fiber are precisely adjusted to maximize the conversion efficiency of the PPLN waveguide. After the conversion, only the signal at 456 nm can be coupled into the fiber through filtering.}
\label{figure2:setup}
\end{figure*}

To overcome the limited range of UV photon transmission from trapped $\text{Yb}^{+}$ ions, a single-stage QFC from UV to telecom band had been indirectly checked by frequency up-conversion of laser light.\cite{JOpt.18.104007, ApplPhysB122.1, PhysRevApplied.7.024021} However, the single-stage process may suffer from large noise due to spontaneous parametric down-conversion (SPDC) and stimulated Raman scattering (SRS). \cite{pelc2012thesis} A two-stage QFC scheme which converts UV photons to NIR range and then to telecom band has also been proposed.\cite{clark2012thesis} However, fabricating a quasi-phase-matched nonlinear crystal for this conversion remains challenging, as it requires an extremely short poling period of less than 2 \unit{\micro\meter} to achieve high conversion efficiency.

For a noiseless, highly efficient conversion of UV photons to the telecom band, we propose to leverage three-stage difference-frequency generation (DFG) process \cite{pelc2012thesis} as shown in Fig. \ref{figure1:convstage}. DFG 1 converts photons from  UV to the visible range, alleviating the requirements for short poling periods and achieving high conversion efficiency. DFG 2 still utilizes a pump laser with a longer wavelength than the output photons at 864 nm to avoid the noise from SPDC. The pump laser for the DFG 2 stage with a wavelength of 965 nm enables the use of the same 1950-nm laser employed in DFG 1 to produce C-band photons in DFG 3. The long wavelength of 1950-nm pump laser used for DFG 1 and 3 also eliminates the SPDC noise. Prior research has demonstrated that DFG stages 2 and 3, operating in a similar wavelength range, can effectively convert visible photons to the telecom band.\cite{ApplPhysB122.1, ACSPhotonics.10.8.2861} Therefore, the initial stage, which transforms UV signals into the visible spectrum, is critical for leveraging Yb$^{+}$ ions in the long-distance quantum network. In this work, we demonstrate a highly efficient, noise-free conversion of UV photons to visible ones, addressing the main bottleneck in the development of scalable quantum networks utilizing $\text{Yb}^{+}$ ions. 

We conduct two experiments to confirm that the QFC process genuinely converted UV photons emitted by a $^{174}$Yb$^{+}$ ion to visible wavelengths. Initially, we measure the second-order correlation of the collected photons after QFC to show their quantum statistical property. Next, we demonstrate that the photons emitted by a single trapped $^{174}$Yb$^{+}$ ion are converted to the visible range by measuring the decay time of the ytterbium ion after QFC. 

Figure \ref{figure2:setup} provides overall schematics of our experimental setup. A single-photon is generated by S-P transition of $ ^{174}$Yb$^{+}$ trapped in a micro-fabricated surface trap\cite{8116614} and is collected by an objective lens with high numerical aperture (NA = 0.6). The magnetic field is set perpendicular to the surface of the trap. By aligning the optical axis of the high NA objective lens parallel to the direction of the magnetic field, we can selectively collect the photons emitted from the $\sigma^{-}$-transition of the ion as shown in Fig. \ref{figure2:setup}(a).\cite{Opt.Express.26.39727, PhysRevA.84.063423} The photon from the $\sigma^{-}$-transition is right-circularly-polarized (RCP) or left-circularly-polarized (LCP) along magnetic axis. This LCP (RCP) photon is converted into horizontal(vertical) polarization using a quarter-wave plate (QWP). The polarizing beam splitter (PBS) filters out the vertically polarized photon, allowing only the horizontally polarized photon to couple into the polarization-maintaining (PM) fiber for transmission to the next setup. During the second-order correlation measurement, single photons were generated via Doppler cooling.\cite{PhysRevLett.58.203} For the decay time measurement, as depicted in Fig. 2(a), a $^{174}$Yb$^{+}$ ion was excited using a vertically polarized single laser pulse, which was reflected by the PBS and directed at the ion. In both experiments, photons spontaneously emitted from the P$_{1/2}$ state via the $\sigma^{-}$-transition were collected through the same objective lens and PBS. The emitted single photon is measured using a photo-multiplier tube (PMT), and the arrival time information was recorded by a field programmable gate array (FPGA), which provided a time resolution of 1.25 ns.

The single photons coupled to the PM fiber are fed into the QFC setup. The input photons at 369 nm and the continuous-wave pump laser at 1950 nm pass through separate variable zoom beam expanders (ZBEs) to optimize the conversion efficiency. Both lights are combined by a dichroic mirror (DM) and coupled to a ridge waveguide on a periodically poled lithium niobate (PPLN) with a poling period of 2.9 \unit{\micro\meter}. To couple both lights to the waveguide, we used an objective lens with a focal shift of approximately 100 \unit{\micro \meter} between the wavelengths of 369 nm and 1950 nm. This chromatic aberration of the focusing lens is pre-compensated by adjusting the magnification and collimation of ZBE. After the waveguide, 369-nm signals are eliminated by 456-nm bandpass filters and most of the pump laser is filtered out by the dichroic mirror. The aspheric lens that collimates the output signals of the PPLN waveguide has a large chromatic focal shift and therefore the rest of pump laser is not coupled to the fiber connected to PMT due to its diverging property. 
When there are no ions trapped in the vacuum chamber, we observe no difference in the PMT count rate whether the pump laser is injected or not. The PMT count rate remained between 10-20 Hz for both cases.
The temperature of the PPLN waveguide is maintained using a thermo-electric cooler to satisfy the phase matching condition. The input polarization of the signal and pump lights are matched to the extraordinary axis of the PPLN waveguide.

\begin{figure}[b]
\includegraphics[width=\linewidth]{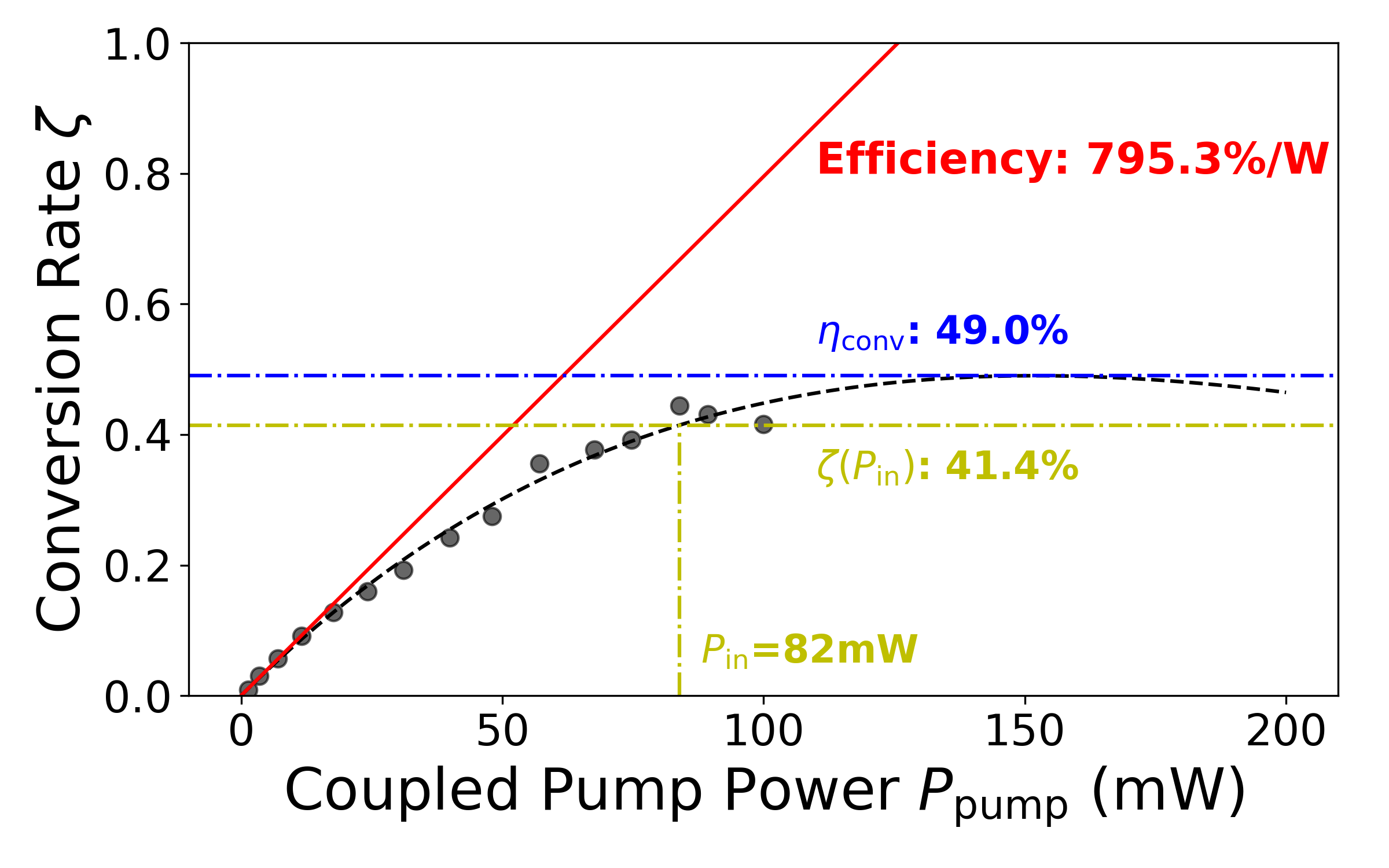}
\caption{Conversion rate of QFC setup. The conversion rate is measured with a low-power 369-nm laser source. Red solid line presents conversion efficiency of 795.3\%/W estimated at low-pump laser power. Black dashed line represents fit to Eq. (\ref{eq:1}), which provides the peak conversion rate $\eta_{\rm conv}$ of 49\% at 152 mW pump laser power. For the safe operation of QFC setup during our experiments, we utilized approximately 82 mW of input pump power $P_{\rm pump}$.}
\label{figure3:freqconv}
\end{figure}

By precise alignment of the ZBE, approximately 20\% of laser light at 369 nm and 50\% of the pump laser were coupled into the PPLN waveguide. The DFG process in the PPLN waveguide achieved a conversion rate of up to 41.4\% for converting a coupled 369-nm laser into 456 nm as shown in Fig. \ref{figure3:freqconv}. The conversion rate $\zeta$ for a given pump power $P_{\rm pump}$ can be described as:
\begin{equation}
\zeta(P_{\rm pump}) = \frac{P_{\rm 456 nm}}{P_{\rm 369 nm}} = \eta_{\rm conv} \sin^{2}\left[(\pi/2) \sqrt{P_{\rm pump}/P_{\rm max}}\right]
\label{eq:1}
\end{equation}
Here, $P_{\rm pump}$ denotes the coupled pump laser power, and $P_{\rm max}$ represents the pump power required to reach the peak conversion rate. The parameter $\eta_{\rm conv}$ represents the maximum achievable conversion rate of the PPLN waveguide, which is constrained by the challenge of coupling the 369-nm signal into the specific spatial mode of the waveguide required for the nonlinear frequency conversion process.

While the PPLN waveguide supports single-mode propagation at 1950 nm, it exhibits highly multi-mode behavior at 369 nm due to the large normalized frequency,\cite{okamoto2010fundamentals} which is primarily attributed to the short wavelength of 369 nm. The objective lens used to focus both lasers into the waveguide exhibits a focal shift of 100 µm. This shift hinders the efficient coupling of the 369-nm laser into the specific spatial mode of the waveguide, while maintaining high coupling efficiency for the 1950 nm laser in the same waveguide. To compensate the focal shift and achieving higher maximum conversion efficiency, $\eta_{\rm conv}$, the ZBE is utilized. Additionally, slight temperature variations or small vibrations in the QFC setup can induce instability in the optical mounts, significantly reducing the coupling efficiency of the 369 nm signal into the target spatial mode of the waveguide. To address these problems, the QFC setup is built on a single water-cooled breadboard mounted on an optical table and enclosed within a protective cover. Under these conditions, the QFC setup maintains more than 70\% of the initial conversion efficiency over a period of 10 hours.

Fitting the measured conversion rate under specific pump power with Eq. (\ref{eq:1}) yielded $P_{\rm max}=\rm 152 \; mW$ and $\eta_{\rm conv}=0.49$ as shown in Fig. \ref{figure3:freqconv}. For the safe operation of QFC over tens of hours, we used about $P_{\rm pump}= \rm 82 \; mW$ of pump power. Note that the conversion rate at single photon level can be described as:
\begin{align*}
\zeta_{\rm photon} = \frac{n_{\rm 456 \; nm}}{n_{\rm 369 \; nm}} = \frac{P_{\rm 456 \; nm} / \hbar \omega_{\rm 456 \; nm}}{P_{\rm 369 \; nm} / \hbar \omega_{\rm 369 \; nm}} = \frac{\omega_{\rm 369 \; nm}}{\omega_{\rm 456 nm}} \times \zeta(P_{\rm pump})
\end{align*}
Therefore, the conversion rate $\zeta(P_{\rm pump})=0.414$ for the given pump power $P_{\rm pump}=82 \rm mW$ corresponds to $\zeta_{\rm photon}=0.512$ at single photon level. More details regarding the measurements of conversion rate can be found in the supplementary material.

The overall conversion efficiency of the QFC setup in Fig. \ref{figure2:setup}(d) includes the transmission of the focusing lens $\eta_{\rm trans}$ and the coupling efficiency of 456-nm light into an optical fiber $\eta_{\rm coupling}^{(\rm fiber)}$. 
\begin{align*}
\eta_{\rm QFC}=\eta_{\rm trans} \; \eta_{\rm coupling}^{(\rm wg)} \; \zeta_{\rm photon} \; \eta_{\rm coupling}^{(\rm fiber)}
\end{align*}
Here, $\eta_{\rm coupling}^{(\rm wg)}$ is the coupling efficiency of 369-nm laser into the waveguide. Note that $\eta_{\rm coupling}^{(\rm wg)} \eta_{\rm conv}$ is proportional to the coupling efficiency into the target spatial mode of the waveguide. According to our measurements, we find $\eta_{\rm coupling}^{(\rm fiber)}=0.7$ and $\eta_{\rm trans}=0.2$. By combining all these values, we have the overall single photon conversion efficiency of our QFC setup.
\begin{align*}
\eta_{\rm QFC}=0.2 \times 0.2 \times 0.512 \times 0.7 = 0.01434
\end{align*}
Additional fiber mating between the single photon generation setup in Fig. \ref{figure2:setup}(a) and $g^{(2)}(0)$ measurement setup in Fig. \ref{figure2:setup}(b) or the QFC setup in Fig. \ref{figure2:setup}(c) degraded the conversion efficiency and we estimate around 1\% of single-photon conversion rate during all of our experiments.

Notably, the low transmission of the focusing lens $\eta_{\rm trans}$ and the poor coupling efficiency of the waveguide $\eta_{\rm coupling}^{(\rm wg)}$ at 369 nm significantly reduce the overall conversion efficiency $\eta_{\rm QFC}$ from 0.3584 to 0.01434. This degradation fundamentally arises from the large wavelength difference between the 369 nm and 1950 nm signals, which makes it challenging for a single lens to achieve high transmittance for both wavelengths and efficiently couple them into a single waveguide. To address this issue, a customized lens with minimal focal shift and high transmittance for both wavelengths can be utilized. The reduced focal shift can facilitate coupling of the 369 nm signal into the target spatial mode of the waveguide, resulting in a higher internal conversion efficiency $\zeta_{\rm photon}$.

\begin{figure}[t]
\includegraphics[width=\linewidth]{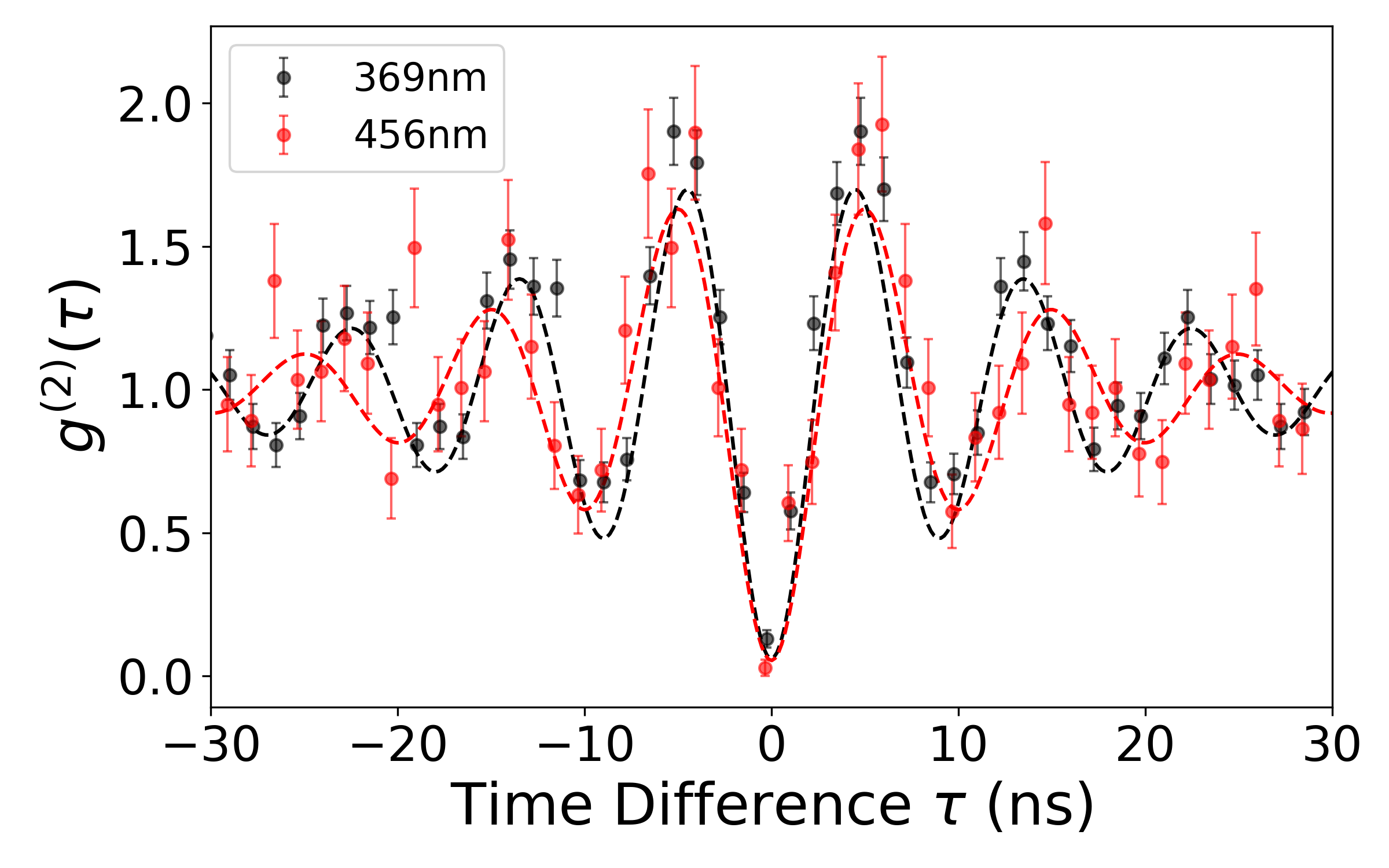}
\caption{Results of the second-order correlation function measurements. Black dots represent the correlation between two photons split by the BS without performing QFC. Red dots show the result of sending one of the photons through the QFC stage for frequency conversion, as shown in Fig~\ref{figure2:setup}(b) and (d) and measuring the correlation. Dotted lines are fitted functions using Eq. (\ref{eq:2}). The measured minimum value of the second-order correlation function is $\text{g}^{(2)}_{\rm min}=0.13 \pm 0.03$ before the conversion and $\text{g}^{(2)}_{\rm min}=0.03 \pm 0.03$ after the conversion. Error bars are standard error assuming Poisson distribution. The fitted parameters give $\text{g}^{(2)}(0)= 0.06$ before conversion and $\text{g}^{(2)}(0)= 0.054$ after conversion.}
\label{figure4:g2}
\end{figure}

We experimentally demonstrated that the converted photon retains its quantum statistics through the $g^{(2)}(0)$ obtained by measuring the time difference of photons detected by two PMTs.\cite{PhysRevLett.58.203} Photons generated during the Doppler cooling process were collected by the PM fiber and subsequently divided by a 50:50 beam splitter (BS) as shown in Fig. 2(b), corresponding to a Hanbury Brown and Twiss (HBT)-type setup. One arm sends photons directly to the PMT, while photons in the other arm undergo QFC, as shown in Fig. ~\ref{figure2:setup}(d) before being detected by the PMT. To compare the effect of QFC, we also sent both photons directly to the PMTs without any QFC. 

The black dots in Fig. ~\ref{figure4:g2} represent the intensity correlation measured for two unconverted photons, while the red dots show the second-order correlation measured when the photon in one arm was converted. $\tau$ represents the time interval between detections by the two PMTs, adjusted for an additional time shift due to the path difference introduced by the QFC stage when the photon in one arm underwent QFC. The dashed lines are fitted using the equation\cite{PhysRevLett.58.203}:
\begin{equation}
g^{(2)}(\tau)=1- g_0 e^{-3\gamma \tau /4}[\cos \Omega \tau+(3\gamma /4\Omega)\sin\Omega \tau],
\label{eq:2}
\end{equation}
where $g_0$ indicates the size of the coincidence dip and $\gamma$ is the natural line width of the relevant energy level of the trapped ion. The oscillation frequency $\Omega$ is defined as:
\begin{align*}
\Omega^2=\Omega_{R}^2+\Delta^{2}-(\gamma/4)^2 
\end{align*}
where $\Omega_{R}$ is the Rabi frequency and $\Delta$ is detuning from the resonance. When fitting the measurement results with Eq. (\ref{eq:2}), we replaced $\tau$ with $\tau - \tau_0$ and set $\tau_0$ as a fitting parameter to account for the limited time resolution of our time stamper (FPGA). The parameters $g_0$, $\gamma$, and $\Omega$ are also used as fitting parameters in Fig. \ref{figure4:g2}. Both graphs exhibit anti-bunching at $\tau = \tau_0$ and the fitted curve yields similar values of $\text{g}^{(2)}(0)$ with and without conversion, resulting in $\text{g}^{(2)}(0) = 0.06$ and $\text{g}^{(2)}(0) = 0.054$, respectively. This indicates that no coincidence events occurred, meaning that the photon emitted from the ion is indivisible. 

\begin{figure}[t]
\centering
\includegraphics[width=\linewidth]{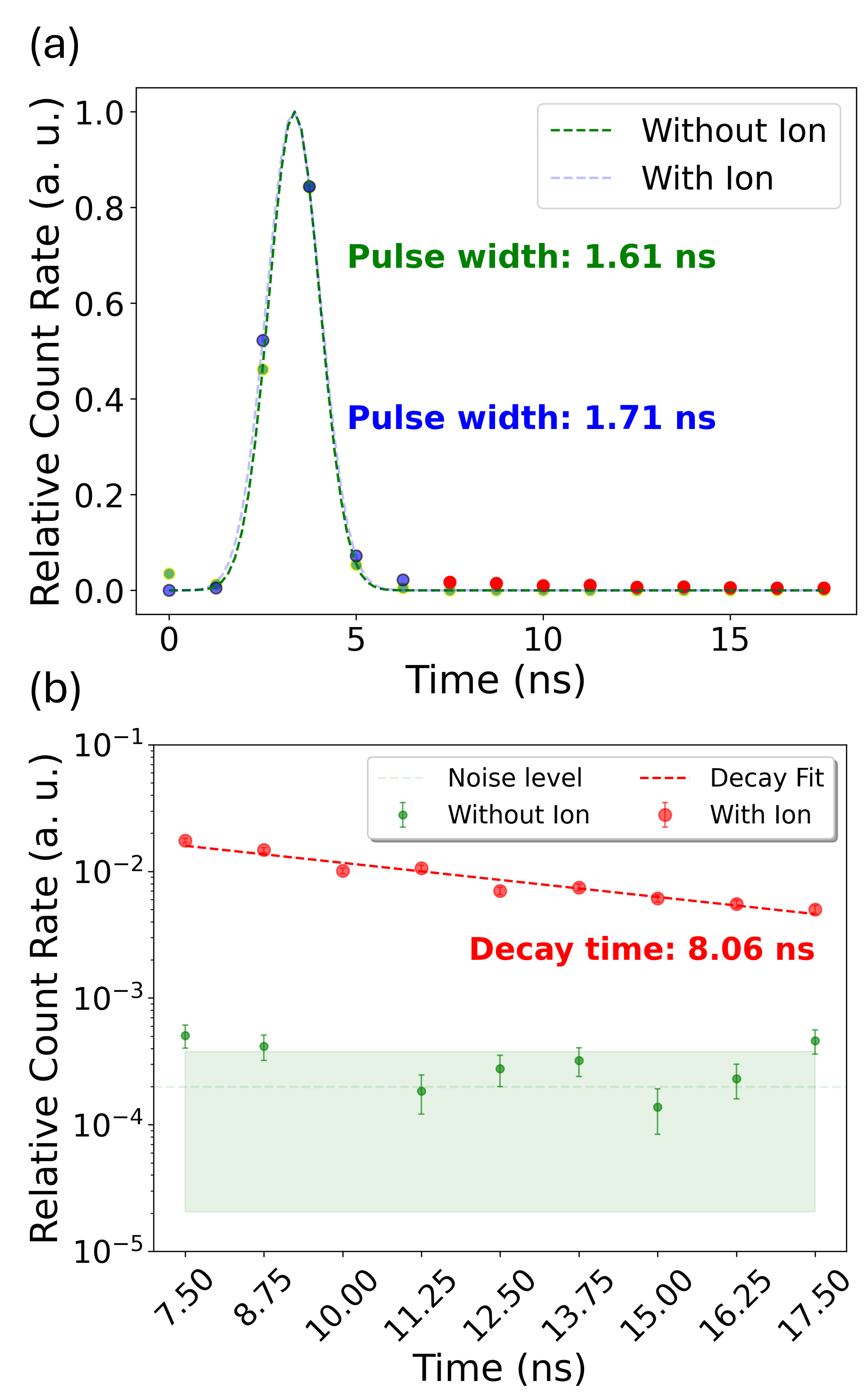}
\caption{Photon arrival time histogram with an ion (represented by blue and red dots) and without an ion (represented by green dots). The vertical axis displays the relative count rate, normalized so that the maxima of the Gaussian fittings are identical to facilitate comparison between the two cases. (a) Photon arrival time histogram including frequency-converted photons from backscattering. Red dots indicate the signals after the 7.5 ns cut-off time. (b) Histogram after the 7.5 ns cut-off time, displayed with a magnified logarithmic vertical scale to highlight the differences. The red dashed line represents a fit from an exponential function, indicating an 8.06 ns decay time of the ion. Note that in the absence of the ion, the signal falls within the green-shaded region, which corresponds to the $\text{1} \sigma$-range of the noise distribution.
}
\label{figure5:decay}
\end{figure}

In the second experiment, the green fibers in Fig. \ref{figure2:setup}(a) are used to excite the $^{174}$Yb$^{+}$ ion to the $\text{P}_{1/2}$ state using a 3-ps pulse laser, and to collect a single photon emitted by the $^{174}$Yb$^{+}$ ion. After converting the wavelength of the collected photons, its arrival time is recorded by FPGA. By measuring the time constant of its exponential decay and comparing this to the natural lifetime of $^{174}$Yb$^{+}$, we can confirm that the single photon we measured indeed originates from $^{174}$Yb$^{+}$.

We extract a single pulse from the mode-locked laser using a pulse picker. The pulse width measured by auto-correlation is 2.79 ps. It is much shorter than the natural lifetime of the $\text{P}_{1/2}$ state, which makes the probability of double excitation negligible. This single-pulse laser is frequency-doubled using a Bismuth Borate (BiBO) crystal to match the S-P transition frequency. The pulse power can be adjusted for optimal excitation using an acousto-optic modulator (AOM).

To measure the time interval between ion excitation and the detection by PMT, the ion is initially Doppler-cooled for 1 \unit{\micro \second}, after which the cooling laser is turned off and a pause of another 1 \unit{\micro \second} follows. Subsequently, the single-pulse laser passes through a PBS and a QWP, becoming RCP and inducing the $\sigma^{-}$ transition. After the ion is excited to the $\text{P}_{1/2}$ state within a few picoseconds, it starts to decay to its ground state, emitting a photon. This photon, after passing through a  QWP and PBS, is coupled into a PM fiber and undergoes QFC process. Following conversion, the photon traverses a filter, and its arrival time at the PMT is measured.

Due to the geometry of the ion trap structure and the inherent limitations of the anti-reflection coating on the viewport, which allows a finite amount of reflection despite the coating, the pulse laser used to excite the ion is partially backscattered and can be coupled into the PM fiber. However, this noise can be distinguished by the arrival times and easily filtered out; the backscattered light arrives immediately after the pulse laser, whereas the emission from $^{174}\text{Yb}^+$ ions adheres to the exponential decay characteristic of an 8.12 ns lifetime. To analyze this, we measure the frequency-converted signals with and without the ion. When the ion is absent, only the backscattered noise would be frequency-converted. Conversely, when the ion is present, both the backscattered noise and the ion's fluorescence would be detected.

Figure \ref{figure5:decay}(a) displays the distribution of arrival times relative to the excitation laser pulse, with the dashed lines representing Gaussian fittings of the backscattered noise in both scenarios - with and without the ion present. The pulse width of a reflected signal is defined as the full width at half maximum (FWHM) of pulse. While the fitted width for the case with the ion is slightly wider than that without the ion, primarily due to the long tail of the exponential decay, the widths in both cases are otherwise almost identical. In Fig. \ref{figure5:decay}(b), the green-shaded region indicates the $1\sigma $-range of the noise distribution, corresponding to the dark count measured when neither the pulse laser nor the ion is present. While the noise level cannot be clearly distinguished from the dark count noise, future works on 3-stage conversion will require more precise characterization of noise sources to bound them from the experimental setup. The red dots represent the photon count with the ion. Therefore, to filter out photons from backscattering, we discard those emitted from $^{174}\text{Yb}^+$ ions before the 7.5 ns cut-off time. Although this approach excludes 42 \% of the total signal, it is important to note that more than half of the signal is still retained. In this result, we obtained the time constant of 8.06 $\pm \text{ 0.9 ns}$, which agrees well with the natural lifetime of $^{174}\text{Yb}^+$ ions at 8.12 ns.\cite{PhysRevA.80.022502}

In conclusion, we demonstrated highly efficient conversion of single UV photons into a visible range. We also verified that converted single photons are truly scattered from $^{174}$Yb$^{+}$ preserving their quantum statistics devoid of any noise. Building upon these results, by integrating the three-stage QFC scheme we proposed, long-distance entanglement can be generated efficiently between two ytterbium ions. Future work will address several technical challenges in maintaining the characteristics of emitted photons across the three conversion stages, including stabilizing the pump laser wavelength and the temperature of the PPLN waveguides. This work presents an experimental approach towards realizing a scalable quantum network based on Yb$^{+}$ ions.

\begin{acknowledgments}
This work was supported by the National Research Foundation of Korea (NRF) grant (No. 2020R1A2C3005689 \& No. RS-2024-00442855) and Institute for Information \& communications Technology Planning \& Evaluation (IITP) grant (No.2022-0-01040, IITP-2024-2021-0-01810), both funded by the Korean government (MSIT). We also thank Dr. Junho Jeong for his programming of the hardware controller, and Minjae Kim and Jiwon Seok for their assistance in coupling the ion fluorescence to the fiber. 
\end{acknowledgments}

\section*{SUPPLEMENTARY MATERIAL}
See the supplementary material for more details on the experimental process regarding the measurement of conversion rate in Fig. \ref{figure3:freqconv}.

\section*{AUTHOR DECLARATIONS}
\subsection*{Conflict of Interest}
The authors have no conflicts to disclose.
\subsection*{Author Contributions}
Seungwoo Yu and Kyungmin Lee contributed equally to this work. \\
\textbf{Seungwoo Yu}: Conceptualization (equal); Formal analysis (equal); Investigation (equal); Software (equal); Methodology (equal); Validation (equal); Visualization (equal); Writing - original draft (equal); Writing - review \& editing (equal). 
\textbf{Kyungmin Lee}: Conceptualization (equal); Formal analysis (equal); Investigation (equal); Software (equal); Methodology (equal); Validation (equal); Visualization (equal); Writing - original draft (equal); Writing - review \& editing (equal). 
\textbf{Sumin Park}: Investigation (supporting); Software (supporting); Methodology (supporting); Validation (supporting); Writing - review \& editing (equal). 
\textbf{Kyunghye Kim}: Methodology (supporting); Software (supporting).
\textbf{Junhong Goo}: Investigation (supporting); Methodology (supporting).
\textbf{Jeonghyun Park}: Software (supporting); Methodology (supporting).
\textbf{Taehyun Kim}: Conceptualization (equal); Validation (equal); Formal analysis (equal); Methodology (equal); Project administration; Resources; Funding acquisition; Supervision; Writing - review \& editing (equal).

\section*{Data Availability Statement}
The data that support the findings of this study are available
from the corresponding author upon reasonable request.

\bibliography{main}
\bibliographystyle{main}

\end{document}



\title{Supplementary Material: Efficient Quantum Frequency Conversion of Ultra-Violet Single Photons from a Trapped Ytterbium Ion}

\author{Seungwoo Yu}
\altaffiliation{These authors contributed equally to this work.}
\affiliation{ 
Department of Computer Science and Engineering, Seoul National University, Seoul 08826, Republic of Korea
}
\affiliation{
Automation and System Research Institute, Seoul National University, Seoul 08826, Republic of Korea}
\affiliation{
Inter-University Semiconductor Research Center, Seoul National University, Seoul 08826, Republic of Korea}
\affiliation{
NextQuantum, Seoul National University, Seoul 08826, Republic of Korea}

\author{Kyungmin Lee}
\altaffiliation{These authors contributed equally to this work.}
\affiliation{ 
Department of Computer Science and Engineering, Seoul National University, Seoul 08826, Republic of Korea
}
\affiliation{
Automation and System Research Institute, Seoul National University, Seoul 08826, Republic of Korea}
\affiliation{
NextQuantum, Seoul National University, Seoul 08826, Republic of Korea}

\author{Sumin Park}
\affiliation{
Automation and System Research Institute, Seoul National University, Seoul 08826, Republic of Korea}
\affiliation{
NextQuantum, Seoul National University, Seoul 08826, Republic of Korea}
\affiliation{ 
Department of Physics and Astronomy, Seoul National University, Seoul 08826, Republic of Korea
}

\author{Kyunghye Kim}
\affiliation{ 
Department of Computer Science and Engineering, Seoul National University, Seoul 08826, Republic of Korea
}
\affiliation{
Automation and System Research Institute, Seoul National University, Seoul 08826, Republic of Korea}
\affiliation{
NextQuantum, Seoul National University, Seoul 08826, Republic of Korea}

\author{Junhong Goo}
\altaffiliation{Currently working at Samsung Display Co., Ltd.}
\affiliation{ 
Department of Computer Science and Engineering, Seoul National University, Seoul 08826, Republic of Korea
}
\affiliation{
Automation and System Research Institute, Seoul National University, Seoul 08826, Republic of Korea}

\author{Jeonghyun Park}
\affiliation{ 
Department of Computer Science and Engineering, Seoul National University, Seoul 08826, Republic of Korea
}
\affiliation{
Automation and System Research Institute, Seoul National University, Seoul 08826, Republic of Korea}
\affiliation{
NextQuantum, Seoul National University, Seoul 08826, Republic of Korea}

\author{Taehyun Kim*}
\email[Author to whom correspondence should be addressed:]{taehyun@snu.ac.kr}
\affiliation{ 
Department of Computer Science and Engineering, Seoul National University, Seoul 08826, Republic of Korea
}
\affiliation{
Automation and System Research Institute, Seoul National University, Seoul 08826, Republic of Korea}
\affiliation{
Inter-University Semiconductor Research Center, Seoul National University, Seoul 08826, Republic of Korea}
\affiliation{
NextQuantum, Seoul National University, Seoul 08826, Republic of Korea}
\affiliation{
Institute of Computer Technology, Seoul National University, Seoul 08826, Republic of Korea}
\affiliation{
Institute of Applied Physics, Seoul National University, Seoul 08826, Republic of Korea}

\maketitle

\section{Measurement of Conversion Rate}
\renewcommand{\thefigure}{S\arabic{figure}}
\begin{figure}[b]
\includegraphics[width=0.8\linewidth]{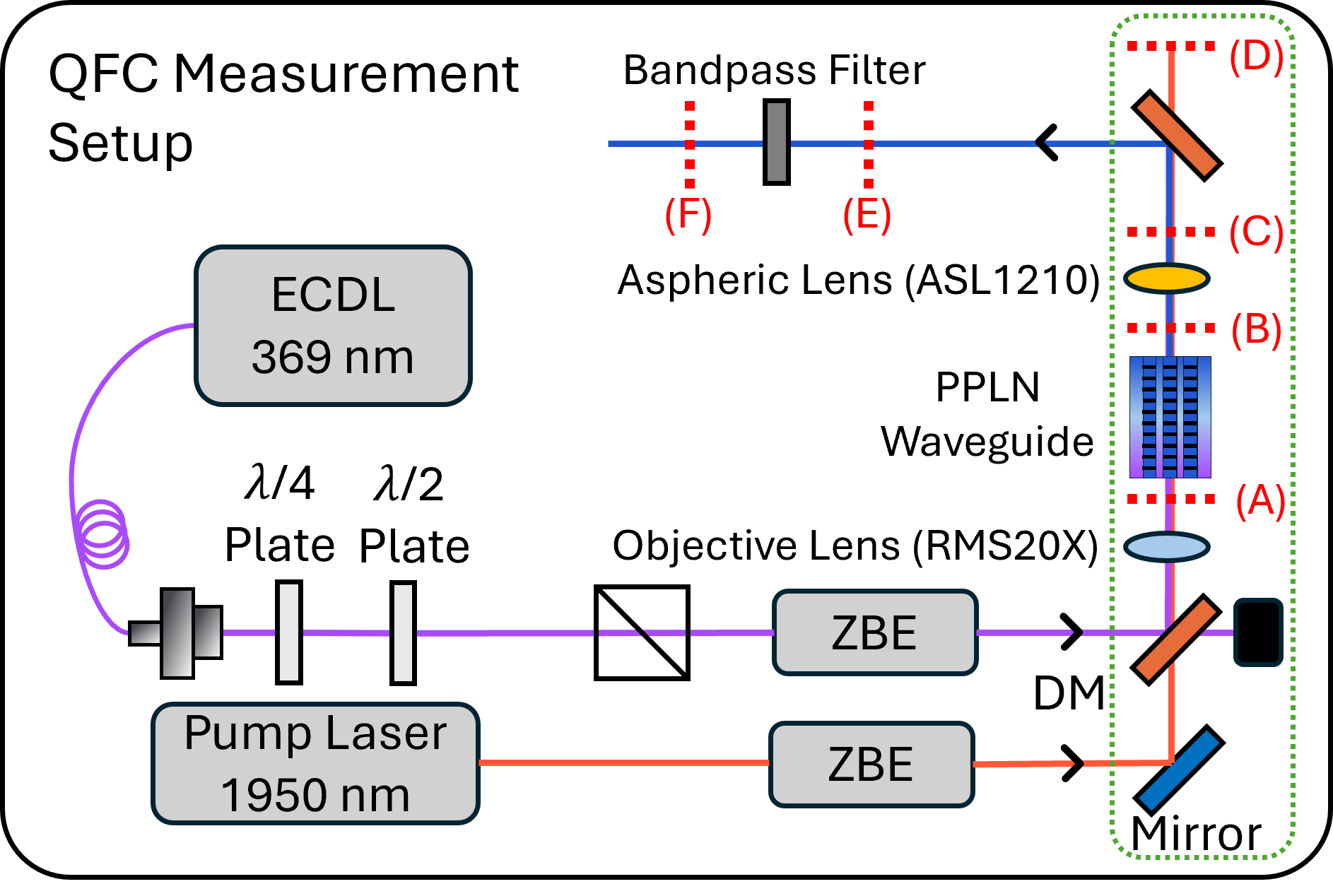}
\caption{The schematic of frequency conversion setup for measuring the conversion efficiency. Continuous-wave lasers at 369 nm and 1950 nm were used to measure the conversion rate. The whole setup was installed in a single water-cooled breadboard. Due to the limited space for measuring pump laser powers in regions (A), (B), and (C), we recorded reference values at low pump power input and used them to estimate the coupled pump laser power. After measuring the pump laser power, we measured the conversion rate by monitoring the coupled 369 nm power in region (E) without pump laser and converted 456 nm power after filtering at location (F).}
\label{figure_s1:freqconv}
\end{figure}

In this section, we provide detailed explanations regarding our experiments to measure the conversion efficiency of the quantum frequency conversion (QFC) setup. Figure \ref{figure_s1:freqconv} illustrates the overall QFC setup. We used the Olympus plan achromat objective lens (RMS20X) due to its minimal focal shift at the time of the experiment. The focal shift between 369 nm and 1950 nm was measured by coupling continuous-wave (CW) lasers at both wavelengths into a single optical fiber and was determined to be approximately 100 \unit{\micro\meter}. The transmittance of the objective lens at 369 nm was evaluated using a CW laser, with about 20\% of the incident 369-nm light being transmitted. For the 1950 nm pump laser, the transmittance was measured to be approximately 50\%. However, this did not affect the conversion efficiency due to the availability of sufficient pump laser power.

Since the entire setup was constructed on a single water-cooled breadboard, we had a limited space for positioning the mirror, dichroic mirror, objective lens, PPLN waveguide, aspheric lens, and another dichroic mirror in a straight line, as shown in the green dashed box in Fig. \ref{figure_s1:freqconv}. This restricted space made it impossible to measure the output power of the PPLN waveguide in region (C) of Fig. \ref{figure_s1:freqconv} after constructing the QFC setup. Also, due to the short focal lengths of the objective lens and aspheric lens, it was not possible to measure the powers at locations (A) and (B) in Fig. \ref{figure_s1:freqconv}. 

To decide the 1950 nm pump laser power coupled to the PPLN waveguide, we recorded reference values at each region in Fig. \ref{figure_s1:freqconv} under low 1950 nm pump power input. For 0.78\unit{\milli\watt} input at (A), $P_{\rm A} = 0.78 \unit{\milli\watt}$, we measured $P_{\rm C}=0.42\unit{\milli\watt}$, $P_{\rm D}=0.3\unit{\milli\watt}$. We observed that the power at (B) was measured to be the same as $P_{\rm C}$ due to the highly divergent output signal from the PPLN waveguide and the inhomogeneous spatial sensitivity of our sensor. However, it does not significantly affect the measured value of the conversion efficiency since the transmittance of the aspheric lens at 1950 nm, which is approximately 0.94, is high enough. Therefore, we used $P_{\rm C}$ as the 1950 nm pump power coupled to the waveguide. Using these values, we determined the ratio $\eta_{\rm r}$ of pump power between (C) and (D):
$$\eta_{\rm r} = \frac{P_{\rm D}}{P_{\rm C}} = 0.71$$
and we also recorded the coupling efficiency $\eta_{\rm pump}^{\rm (ref)}$ of the PPLN waveguide for 1950-nm input:
$$\eta_{\rm pump}^{\rm (ref)} = 0.54,$$
which were used as reference values at high pump power input. 
\renewcommand{\thetable}{S\arabic{table}}
\begin{table}[t]
\includegraphics[width=0.9\linewidth]{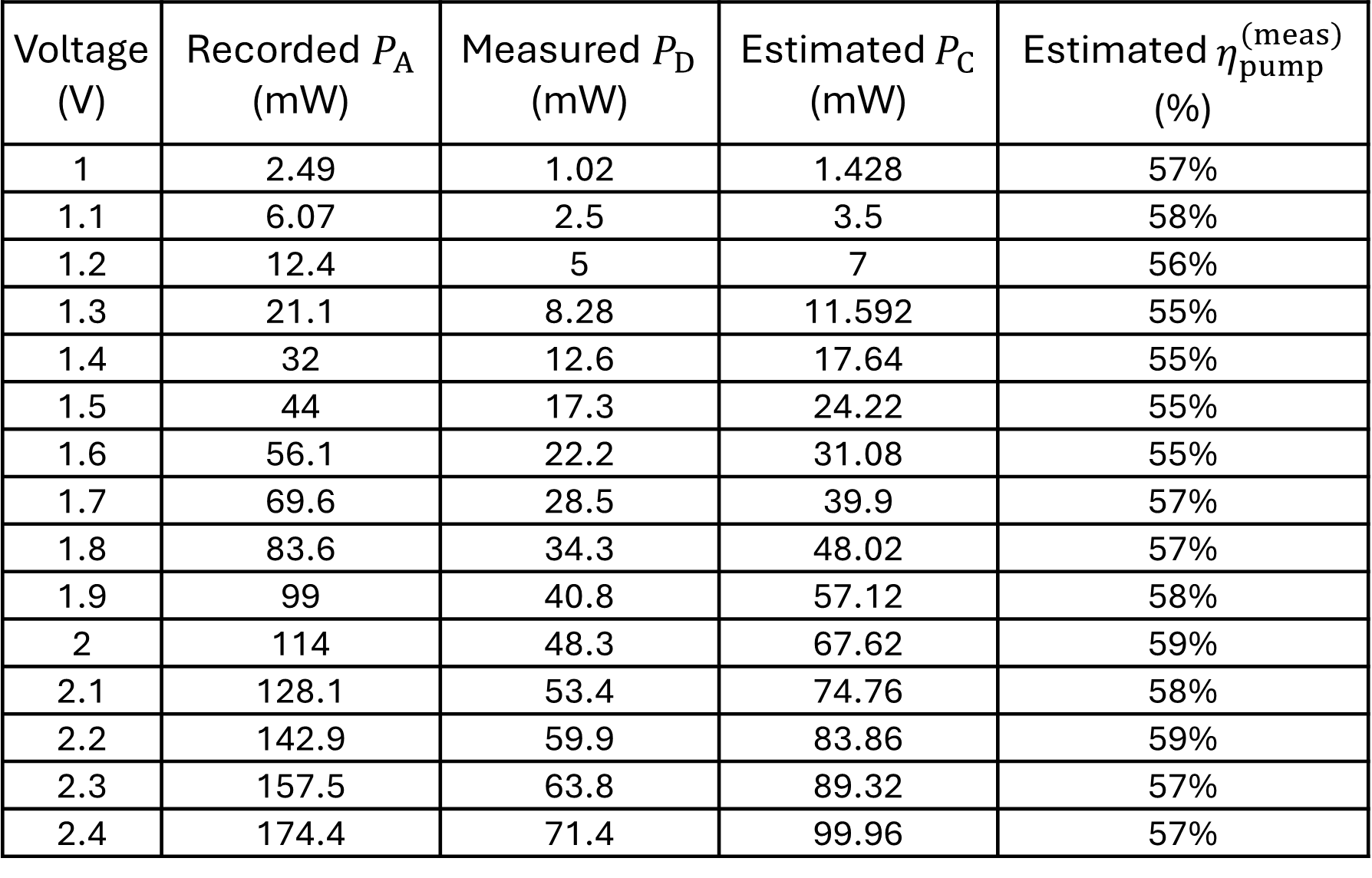}
\caption{Recorded pump laser power at each region in Fig. \ref{figure_s1:freqconv} during the experiment. Voltage shows the applied volatages to the 1950 nm pump amplifier. Recorded powers in region (A), $P_{\rm A}$, were pre-measured before all the optical elements are positioned. $P_{\rm D}$ were measured during the experiments in region (D) and were used to have the estimated power at location (C), $P_{\rm C}$. Using these values, we monitored the estimated coupling efficiency of 1950 nm pump laser. The estimated coupling efficiencies $\eta_{\rm pump}^{\rm (meas)}$ were similar to the reference value $\eta_{\rm pump}^{\rm (ref)}=54\%$.}
\label{table_s1:pumpmeasure}
\end{table}

The input pump power can be controlled by adjusting the voltage applied to the 1950 nm amplifier. For each voltage input, we pre-measured the pump power at location (A) $P_{\rm A}^{\rm (ref)}$ in Fig. \ref{figure_s1:freqconv}. After finishing the recording, we placed the waveguide and coupled the pump laser into the waveguide. The coupled 1950 nm pump power $P_{\rm D}$ at location (D) in Fig. \ref{figure_s1:freqconv} was monitored during the experiment. Using $\eta_{\rm pump}^{\rm (meas)}$ and $\eta_{\rm r}$, we could estimate pump powers in region (C) from the measured value of $P_{\rm D}$ as below:
$$P_{\rm C} = \frac{P_{\rm D}}{\eta_{\rm r}}$$
Then we estimated the coupling efficiency using the pre-measured input pump power $P_{\rm A}^{\rm (rec)}$:
$$\eta_{\rm pump}^{\rm (meas)} = \frac{P_{\rm C}}{P_{\rm A}^{\rm (rec)}}.$$
We monitored the estimated coupling efficiency $\eta_{\rm pump}^{\rm (meas)}$ with the reference value $\eta_{\rm r}$ during the experiment as shown in table \ref{table_s1:pumpmeasure}. The estimated coupling efficiencies were similar to the reference value $\eta_{\rm pump}^{\rm (ref)}=54\%$, indicating that the coupled 1950 nm pump laser power had not varied significantly.

\begin{table}
\includegraphics[width=0.9\linewidth]{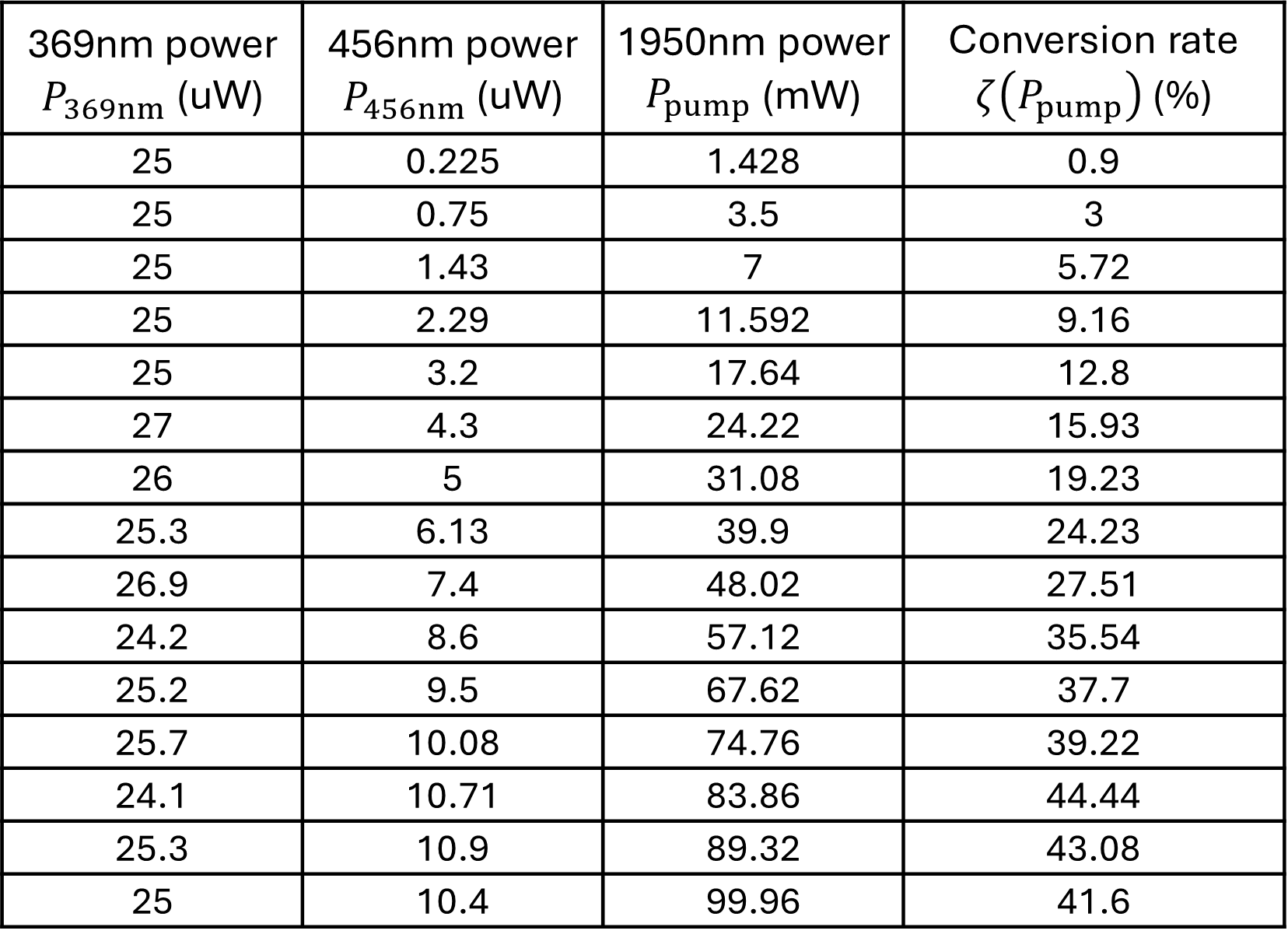}
\caption{Measurement results of conversion rate under various input pump power. The coupled 369 nm power was measured in region (E) of Fig. \ref{figure_s1:freqconv} and converted 456 nm signal was detected at location (F). The 1950 nm pump laser power $P_{\rm pump}$ was estimated by using the observed pump power $P_{\rm D}$ in region (D).}
\label{table_s2:convmeas}
\end{table}

After checking the stability of the 1950 nm pump laser coupling efficiency $\eta_{\rm pump}^{\rm (meas)}$, we measured the conversion rate as shown in table \ref{table_s2:convmeas}. Before applying 1950 nm pump laser power, we measured the coupled 369 nm power without pump laser input in region (E) of Fig. \ref{figure_s1:freqconv} and observed about 25 \unit{\micro\watt} output at 369 nm from the PPLN waveguide. We also confirmed that no 369 nm power was detected in region (F) of Fig. \ref{figure_s1:freqconv} after passing through the bandpass filter, with a measurement resolution of 1 \unit{\nano\watt}. After turning on the pump laser, we measured the converted 456-nm signal after the bandpass filter. The conversion rate $\zeta (P_{\rm pump})$ is defined simply as the ratio between the converted 456-nm power and coupled 369-nm power under the given pump power:
$$\zeta(P_{\rm pump}) = \frac{P_{\rm 456nm}}{P_{\rm 369nm}}.$$
By fitting the measured conversion rates in Table \ref{table_s2:convmeas} with Eq. (1) in the main text, we obtained the results presented in Fig. 3 of the main text.
